\documentstyle[aps,prb,multicol]{revtex}
\begin{document}
\title{Accurate calculation of polarization-related quantities in
semiconductors}
\author{Fabio Bernardini,$^{1}$ Vincenzo Fiorentini,$^{1,2}$ and
David Vanderbilt$^{3}$}
\address{1) INFM and Dipartimento di Fisica, Universit\`a di Cagliari,
Italy\\ 2) Walter Schottky Institut, TU M\"unchen, Garching, Germany
\\ 3) Department of Physics and Astronomy, Rutgers University,
Piscataway, New Jersey 08845-0849} 
\date{\today}
\maketitle\begin{abstract} We demonstrate that
polarization-related quantities in semiconductors can be
predicted accurately from first-principles calculations using the
appropriate approach to the problem, the Berry-phase polarization
theory. For III-V nitrides, our test case, we find polarizations,
polarization differences between nitride pairs, and piezoelectric
constants quite close to their previously established values.
Refined data are nevertheless provided for all the relevant
quantities.
\end{abstract}

\begin{multicols}{2}

The importance of spontaneous and piezoelectric polarization in
nitride semiconductors is by now widely recognized, and numerous
reports of observations and practical exploitations of polarization
effects in optoelectronic and electronic nitride nanostructure
devices have appeared.\cite{exp} Many experiments have
been rather successfully interpreted using recently
calculated {\it ab initio} values of the spontaneous polarization and the
piezoelectric constants.\cite{noi-prb1}

Doubts have recently been raised about the accuracy of computed
spontaneous polarization values in nitride semiconductors.  A recent
study\cite{bechstedt}  based on a supercell method\cite{posternak}
reported values of the spontaneous polarization
of AlN, GaN, and InN that differ widely (by $\sim$50-100\%)
from those of our recent  calculations.\cite{noi-prb1} Our results
were obtained within density-functional theory using
numerical methods based on the modern theory of polarization and
the Berry-phase concept,\cite{ksv} so they are expected to be
highly reliable. Indeed, they have been used successfully in the
experimental\cite{exp} and modeling\cite{mod} literature. It is
therefore appropriate to reexamine the accuracy of the polarization
calculations, to dissipate doubts and confusion about which values
to use, and to clarify the size of the expected errors.
Since the nitrides provide a severe test for polarization calculations,
we expect our error estimates to provide a valid upper bound
for all semiconductors.

The message of  the present paper is fourfold.  {\it First}, we
find that it  {\it is} possible to calculate polarizations with the
accuracy needed to compare with experiment,\cite{exp,mod} provided
a well-controlled and accurate method is used to calculate the
polarization, and that the structure is accurately optimized by
state-of-the-art {\it ab initio} density-functional calculations.  The
polarization is found to be most sensitive to internal structural
parameters (in wurtzite, the parameter $u$).  {\it Second},
absolute polarization values resulting from the present revision
deviate only slightly ($\sim$10\%) from the previous values.  Moreover,
observable quantities, namely polarization {\it differences}
between various compounds, are essentially unchanged with
respect to our previous report.\cite{noi-prb1} {\it Third}, we
report revised values  of the  piezoelectric constants, again
rather close to our previous data; we consider that the present
values are more refined and should be preferred in modeling. We
accompany the revised piezoconstants by accurately calculated
elastic constants, providing a self-consistent set of data to
estimate piezoelectric fields in strained nitrides. {\it
Fourth},  we briefly address the possible origin of the
discrepancy between the report of Ref.~\onlinecite{bechstedt} and
our results.  We suggest that it may be connected with technical
shortcomings, especially imperfect convergence of the equilibrium
structure, in the superlattice calculation,

We analyze the polarization as obtained using the Berry-phase
method\cite{noi-prb1,ksv} within two different density-functional
theory (DFT) exchange-correlation schemes.  Specifically, as do
the authors of Ref.\onlinecite{bechstedt}, we use the VASP package
\cite{vasp} and the pseudopotentials provided therewith.  We carry
out calculations using both the generalized gradient approximation
(GGA) to density-functional theory in the Perdew-Wang PW91
version, and the local-density approximation (LDA) in the
Ceperley-Alder-Perdew-Zunger form (used in Ref.  \onlinecite{bechstedt}).
Ultrasoft potentials\cite{ultrasoft} are used (Ga and In
$d$ electrons are treated as valence) at a conservative cut-off of 325
eV,  and reciprocal-space summation is done on an (888)
Monkhorst-Pack  mesh.

Our results are as follows. In Tables \ref{tab.gan},
\ref{tab.aln} and \ref{tab.inn}  we list the structural parameters
$a$, $c/a$, and $\epsilon_1 = u-u_{\rm ideal}$, and the spontaneous
polarization $P_{\rm sp}$, for the III-V nitrides. To obtain them,
we optimize the structure within both the GGA and LDA,
then calculate the spontaneous polarization with the Berry-phase
technique. In addition, we calculate the polarization for  sets of
structural parameters reported by others.  Specifically, we use the
LDA-pseudopotential lattice parameters  reported by Bechstedt,
Gro\ss{}ner, and Furthm\"uller \cite{bechstedt} (BGF), and those
calculated with the FLAPW method within the LDA by Wei and
Zunger (WZ).\cite{wei}  The  results  are labeled  `Present (LDA)',
`Present (GGA)', `BGF (LDA)', and `WZ (LDA)' in Tables
\ref{tab.gan} to \ref{tab.inn}. Next,  in Table \ref{tab.dP} we
give the spontaneous polarization {\it differences} for some
relevant cases. Finally, in Table \ref{tab.new} we report the
piezoelectric constants, dynamical Born charges, spontaneous
polarizations, and elastic constants (relevant to the piezofields in
strained nitrides layers) as calculated in the GGA approximation,
which are proposed as revised values to be used in modeling of
experiments involving macroscopic polarization effects.

From Tables \ref{tab.gan} to \ref{tab.inn}, it  is evident that the
Berry-phase results are quite homogeneous despite the
methodological differences of the various methods and DFT
parameterizations (the main exception, GaN in the Bechstedt
structure, is discussed below).  As pointed out below, they differ
considerably from the values calculated \cite{bechstedt} in the
supercell approach.  The major source of these  deviations are
inaccuracies in the determination of the internal parameter $u$. In
the ideal wurtzite structure, $u$=$u_{\rm ideal}$=0.375 in units
of  $c/a$. Here we compare different calculations through the
parameter $\epsilon_1=u - u_{\rm ideal}$, expressed  in units of
10$^{-3}\, c/a$.
It can be noticed from the
Tables that polarization is very sensitive to $\epsilon_1$, though
relatively insensitive to other structural parameters, as
also noted previously.\cite{posternak,posternak2}  For GaN, the
$\epsilon_1$ value of Ref.~\onlinecite{bechstedt} deviates
significantly from the others, and indeed we calculate a
corresponding Berry-phase $P_{\rm sp}$ which is more than a factor of
two larger than all the others.  The polarization of $-$0.074\,C/m$^2$
obtained in the superlattice calculations \cite{bechstedt}
is also too large by a factor 2. This suggests that the origin of
the discrepancy for GaN is the overestimation of the $\epsilon_1$
parameter ($\epsilon_1$=6.5) in Ref.~\onlinecite{bechstedt}. Since
the same group reported\cite{hyper}  $\epsilon_1$=1.5 for GaN (in
good agreement with other entries in our Table and with
experiment) using the same code  and pseudopotentials, but a much
finer k-point summation mesh, the deviation is possibly due to
k-space summation.

For AlN, the largest discrepancy (between the values for the GGA and
LDA--Bechstedt structures) is  13\,mC/m$^2$, i.e., a difference
of $\sim$1.0 in $\epsilon_1$. Indeed, the GGA structure is closer
to experiment, and it should be more reliable. In any event, the
superlattice value of $-$0.120\,C/m$^2$ is about 20-25\,mC/m$^2$, or
25\%, larger than all the Berry-phase values.  For InN,
$\epsilon_1$ is quite homogeneous throughout, and indeed so are the
polarizations, all within 5\%.  The superlattice polarization is
$-$0.050\,C/m$^2$, also in the vicinity of the Berry-phase values.

From the data in  Tables \ref{tab.gan}, \ref{tab.inn}, and
\ref{tab.aln}, it appears that the Berry-phase  polarization values
are quite stable and largely independent of exchange-correlation
scheme,  once the structure (mostly, the $\epsilon_1$ parameter)
is determined correctly.  The polarizations previously
determined\cite{noi-prb1} for AlN, GaN, and InN were $-$0.081,
$-$0.029, $-$0.032; the present GGA values are  11\%,  9\% and  23\%
larger. We also remark that the GGA values obtained in this paper
produce polarization differences between nitride pairs that compare
quite well with those in current use.\cite{noi-prb1}  Table
\ref{tab.dP} reports the comparison.

We also recalculated the dynamical charges and piezoelectric
constants of the nitrides in the GGA approximation to verify our
earlier results.  Also, we evaluated within the GGA the elastic
constants needed in the determination of strain in epitaxial
nitride layers, which bears on their built-in piezoelectric
fields.  The results are fairly close to our previously reported
values (Table II of Ref.~\onlinecite{noi-prb1}). The data
 listed in Table \ref{tab.new} are proposed as updated reference 
values.

It has recently been shown\cite{proper}  that
the proper piezoelectric response is related to a current flow
across a sample in response to polarization. This response
 corresponds to a set of so-called {\it proper} piezoelectric
constants. These  differ  from the polarization derivatives with
respect to strain (the {\it improper} constants) by a quantity
proportional to the spontaneous polarization, hence of  order 10\%.  
The piezoelectric constants just reported in Table \ref{tab.new} are
the improper constants.\cite{proper}   These constants are those to be
used in the interpretation or modeling  of
experiments involving depolarizing fields and polarization-induced
interface charges.\cite{noi-prb2} On the other hand, 
in view of their definition,\cite{proper} the
 proper  constants are those to be compared with
experiments based on measurements of {\it current} across
piezoelectric samples and
with values calculated from linear-response theory.\cite{dfpt-e}
The proper constants as calculated within the GGA
are reported in Table \ref{tab.new2}. Only the $e_{31}$'s are affected,
while the $e_{33}$'s are unchanged.

Going back to the accuracy of calculated polarizations,
it is appropriate to note that the theoretical results are relatively
stable  in other systems also. For BeO (note that Ref.
\onlinecite{bechstedt}  misquotes Refs. \onlinecite{noi-prb1} and
\onlinecite{posternak} about this material), a longitudinal
polarization $P_{\rm L}$= $-$15\,mC/m$^2$ is reported\cite{posternak}
for  a  length-optimized clamped-ion wurtzite half cell. This
corresponds to a transverse $P_{\rm T}$ = $\varepsilon_{\infty}$
$P_{\rm L}$= $-$48\,mC/m$^2$, in excellent  agreement with our
Berry-phase value\cite{noi-prb1}  of $-$50\,mC/m$^2$. Also,  the
transverse  spontaneous polarization  of  ZnO was found to be
$-$57\,mC/m$^2$ using pseudopotentials \cite{noi-prb1} and $-$50\,mC/m$^2$
using FLAPW.\cite{posternak2}    Again, the accuracy depends most
critically on the internal  parameters.  A similar but more
stringent test for GaN/AlN strained superlattices \cite{noi-prb2}
showed that  the calculated interface charge agrees to within less
than 1\,mC/m$^2$ with that predicted from bulk transverse
polarizations and dielectric constants.\cite{noi-prl}

We briefly note that the structural issues  discussed above do not
account for all the discrepancies observed with the calculations of
Ref.\onlinecite{bechstedt}.  Two possible additional sources of error are
related to the presence of large depolarizing  electrostatic fields
in the supercell (see Fig.\,3 of Ref.\onlinecite{bechstedt}).
First, the  length of the clamped-ion superlattice  was not
optimized. In a non-zero field $E_z$, this has the same effect
\cite{noi-prl} of an inverse-piezoelectric distortion, leading
to a spurious polarization $\delta P_z = - [e^{(0)}_{33}]^2\, E_z/C_{33}$
(with $e^{(0)}_{33}$ the clamped-ion piezoconstant and $C_{33}$ the elastic
modulus).  However, this is only a  small
correction ($\sim$1 mC/m$^2$).  Second, the field causes a large
potential drop over the superlattice, amounting respectively to
about 1.4, 3.3 and 6.0 eV  in InN, GaN, and AlN (in the wurtzite
region; see Fig.~3 of Ref.~\onlinecite{bechstedt}).  This drop is
larger than  the calculated  LDA gap of the materials. While there
seems to be no evidence for instability,\cite{bechstedt} we presume
this could be a significant source of error.

In summary we have shown that the polarization of nitride
semiconductors can be calculated within the DFT-based
first-principles Berry-phase approach  in a stable and reproducible
manner. Calculated polarizations can therefore be  used safely in
comparisons with experiment in nitride quantum structures.
Given the demanding test case provided by the nitrides, we believe
this conclusion applies to any semiconductor  whose structure, and
in particular whose internal parameters, can be accurately
predicted.

Work at Cagliari University is supported in part by MURST--Cofin99
and   INFM Parallel Computing Initiative.  V.F. thanks the Alexander
von Humboldt-Stiftung for support of his stays at the Walter
Schottky Institut.
D.V. acknowledges support from NSF grant DMR-9981193.

\narrowtext
\begin{table}
\caption{Structure and polarization of GaN.
The Berry-phase spontaneous polarization $P_{\rm sp}$ has been
calculated for each structure and setting listed. The structures
are those obtained in the LDA by  Bechstedt, Gro\ss{}ner, and
Furthm\"uller \protect\cite{bechstedt} (BGF) and by Wei and
Zunger\protect\cite{wei} (WZ), and by ourselves (Present) in both
the LDA and GGA.  Lattice constant $a$ in \AA, $\epsilon_1$ in
10$^{-3} c/a$, and $P_{\rm sp}$ in C/m$^2$.
Experimental values from Ref.~\protect\onlinecite{schu}.
Ref.~\protect\onlinecite{bechstedt} reports a supercell-based
$P_{\rm sp}$=$-$0.074\,C/m$^2$ for the BGF (LDA) case.
}
\label{tab.gan}
\begin{tabular}{lcccc}
\multicolumn{1}{l}{Structure from} &
\multicolumn{1}{c}{$a$} & 
\multicolumn{1}{c}{$c/a$} & 
\multicolumn{1}{c}{$\epsilon_1$} & 
\multicolumn{1}{c}{$P_{\rm sp}$}\\
\hline
BGF (LDA) & 3.150 & 1.6310 & 6.5  & $-$0.080\\
WZ (LDA) & 3.180 & 1.6259 & 1.8  &  $-$0.032 \\
Present (LDA) & 3.131 & 1.6301  & 1.6 & $-$0.032   \\
Present (GGA) & 3.197 & 1.6257 & 1.9  & $-$0.034 \\
Experiment & 3.19 & 1.627 & 2.0 & --- \\
\end{tabular}
\end{table}

\begin{table}
\caption{Structure and polarization of AlN.
Details as in Table \protect\ref{tab.gan}.
Ref.~\protect\onlinecite{bechstedt} reports a supercell-based
$P_{\rm sp}$=$-$0.120\,C/m$^2$ for the BGF (LDA) case.
}
\label{tab.aln}
\begin{tabular}{lcccc}
\multicolumn{1}{l}{Structure from} &
\multicolumn{1}{c}{$a$} & 
\multicolumn{1}{c}{$c/a$} & 
\multicolumn{1}{c}{$\epsilon_1$} & 
\multicolumn{1}{c}{$P_{\rm sp}$}\\
\hline
LDA (BGF) & 3.080 & 1.6070 & 7.4  & $-$0.103\\
LDA (WZ) & 3.112 & 1.6009 & 6.9  &  $-$0.094 \\
LDA (present) & 3.070 & 1.5997  & 7.1 & $-$0.099   \\
GGA (present) & 3.108 & 1.6033 & 6.4  & $-$0.090 \\
Experiment & 3.11 & 1.601 & 7.1 & --- \\
\end{tabular}
\end{table}

\begin{table}
\caption{Structure and polarization of InN.
Details as in Table \protect\ref{tab.gan}.
Ref.~\protect\onlinecite{bechstedt} reports a supercell-based
$P_{\rm sp}$=$-$0.050\,C/m$^2$ for the BGF (LDA) case.
}
\label{tab.inn}
\begin{tabular}{lcccc}
\multicolumn{1}{l}{Structure from} &
\multicolumn{1}{c}{$a$} & 
\multicolumn{1}{c}{$c/a$} & 
\multicolumn{1}{c}{$\epsilon_1$} & 
\multicolumn{1}{c}{$P_{\rm sp}$}\\
\hline
LDA (BGF) & 3.530 & 1.6320 & 5.0  & $-$0.043\\
LDA (WZ) & 3.544 & 1.6134 & 4.0  &  $-$0.042 \\
LDA (present) & 3.509 & 1.6175  & 3.6 & $-$0.041  \\
GGA (present) & 3.580 & 1.6180 & 3.7  & $-$0.042 \\
Experiment & 3.544 & 1.613 & --- & --- \\
\end{tabular}
\end{table}

\begin{table}
\caption{Modulus of the spontaneous-polarization differences
(mC/m$^2$) between binary nitrides. The values of
Ref.~\protect\onlinecite{noi-prb1}  and
Ref.~\protect\onlinecite{bechstedt} are reported 
together with the present GGA result.}
\label{tab.dP}
\begin{tabular}{lccc}
\multicolumn{1}{c}{$\mid\Delta P\mid$} & 
\multicolumn{1}{c}{Present (GGA)}& 
\multicolumn{1}{c}{Ref.\onlinecite{noi-prb1}}& 
\multicolumn{1}{c}{Ref.\onlinecite{bechstedt}} \\
\hline
AlN/GaN & 55 & 52 & 46 \\
GaN/InN & 7 & 3 & 24 \\
InN/AlN & 48 & 49 & 70 \\
\end{tabular}
\end{table}

\begin{table}
\caption{Spontaneous polarization (C/m$^2$),
improper piezoelectric constants (C/m$^2$), 
 dynamical charges, and elastic constants (GPa)
 for the nitrides as obtained in
the GGA approximation.}
\label{tab.new}
\begin{tabular}{lcccccc}
\multicolumn{1}{c}{}&
\multicolumn{1}{c}{$P_{\rm sp}$}&
\multicolumn{1}{c}{$e_{33}$}&
\multicolumn{1}{c}{$e_{31}$}&
\multicolumn{1}{c}{$Z^{*}$} &
\multicolumn{1}{c}{$C_{33}$}&
\multicolumn{1}{c}{$C_{31}$}\\
\hline
AlN & $-$0.090 & 1.50 & $-$0.53 & 2.65 & 377 & 94\\
GaN & $-$0.034 & 0.62 & $-$0.34 & 2.76 & 354 & 68 \\
InN & $-$0.042 & 0.81 & $-$0.41 & 3.10 & 205 & 70 \\
\end{tabular}
\end{table}

\begin{table}
\caption{Proper
piezoelectric constants (C/m$^2$)
 for the nitrides in
the GGA approximation.}
\label{tab.new2}
\begin{tabular}{lcc}
\multicolumn{1}{c}{}&
\multicolumn{1}{c}{$e_{33}$}&
\multicolumn{1}{c}{$e_{31}$}\\
\hline
AlN &   1.50 & $-$0.62 \\
GaN &   0.62 & $-$0.37 \\
InN &   0.81 & $-$0.45 \\
\end{tabular}
\end{table}

\end{multicols} 
\end{document}